\documentclass[twocolumn,superscriptaddress,longbibliography,aps,pra,floatfix]{revtex4-1}

\usepackage{graphicx}
\usepackage{bm}
\usepackage{color}
\usepackage{amsmath}
\usepackage{amssymb}
\usepackage{epstopdf}
\usepackage{gensymb}
\usepackage{float}
\usepackage[urlcolor=blue,colorlinks=true,citecolor=blue,linkcolor=blue,pdfstartview={FitH},bookmarks=false]{hyperref}
\usepackage{xfrac}
\usepackage{amsmath,bm}

\begin{document}

\title{Trimeron-phonon coupling in magnetite}

\author{Przemys\l{}aw Piekarz}
\affiliation{Institute of Nuclear Physics, Polish Academy of Sciences, Radzikowskiego 152, 31-342 Krak\'{o}w, Poland}

\author{Dominik Legut}
\affiliation{IT4Innovations and Nanotechnology Center, VSB-Technical University
of Ostrava, 17.listopadu 2172/15, 708 00 Ostrava-Poruba, Czech Republic}

\author{Edoardo Baldini}
\author{Carina A. Belvin}
\affiliation{Department of Physics, Massachusetts Institute of Technology, Cambridge, Massachusetts 02139, USA}

\author{Tomasz Ko\l{}odziej}
\affiliation{SOLARIS National Synchrotron Radiation Centre, Czerwone Maki 98, 30-392 Krak\'{o}w, Poland}

\author{Wojciech Tabi\'{s}}
\author{Andrzej~Koz\l{}owski}
\author{Zbigniew Kakol}
\author{Zbigniew Tarnawski}
\affiliation{AGH University of Science and Technology, Faculty of Physics and Applied Computer Science, 30-059 Krak\'{o}w, Poland}

\author{Jos\'e Lorenzana}
\affiliation{La Sapienza University,  00-185 Rome, Italy}

\author{Nuh Gedik}
\affiliation{Department of Physics, Massachusetts Institute of Technology, Cambridge, Massachusetts 02139, USA}

\author{Andrzej M. Ole\'{s}}
\affiliation{\mbox{Institute of Theoretical Physics, Jagiellonian University,
             Profesora Stanis\l{}awa \L{}ojasiewicza 11, 30-348 Krak\'ow, Poland}}
\affiliation{Max Planck Institute for Solid State Research, Heisenbergstrasse 1, 70569 Stuttgart, Germany}

\author{J\"{u}rgen M. Honig}
\affiliation{Department of Chemistry, Purdue University, West Lafayette, Indiana 47907-2084, USA}

\author{Krzysztof Parlinski}
\affiliation{Institute of Nuclear Physics, Polish Academy of Sciences, Radzikowskiego 152, 31-342 Krak\'{o}w, Poland}

\begin{abstract}
Using density functional theory, we study the lattice dynamical
properties of magnetite (Fe$_3$O$_4$) in the high-temperature cubic
and low-temperature monoclinic phases. The calculated phonon dispersion
curves and phonon density of states are compared with the available
experimental data obtained by inelastic neutron, inelastic
x-ray, and nuclear inelastic scattering. 
We find a very good agreement between the theoretical and experimental
results for the monoclinic $Cc$ structure revealing the strong coupling
between charge-orbital (trimeron) order and specific phonon modes.
For the cubic phase, clear discrepancies arise which, remarkably, 
can be understood assuming that the strong trimeron-phonon coupling can be 
extended above the Verwey transition, with lattice dynamics influenced
by the short-range trimeron order instead of the average cubic
structure. Our results establish the validity of trimerons (and trimeron-phonon coupling) 
in explaining the physics of magnetite much beyond their original formulation.

\end{abstract}

\date{\today}

\maketitle

\section{Introduction}

\begin{figure}[b!]
\begin{center}
\includegraphics[scale=0.31]{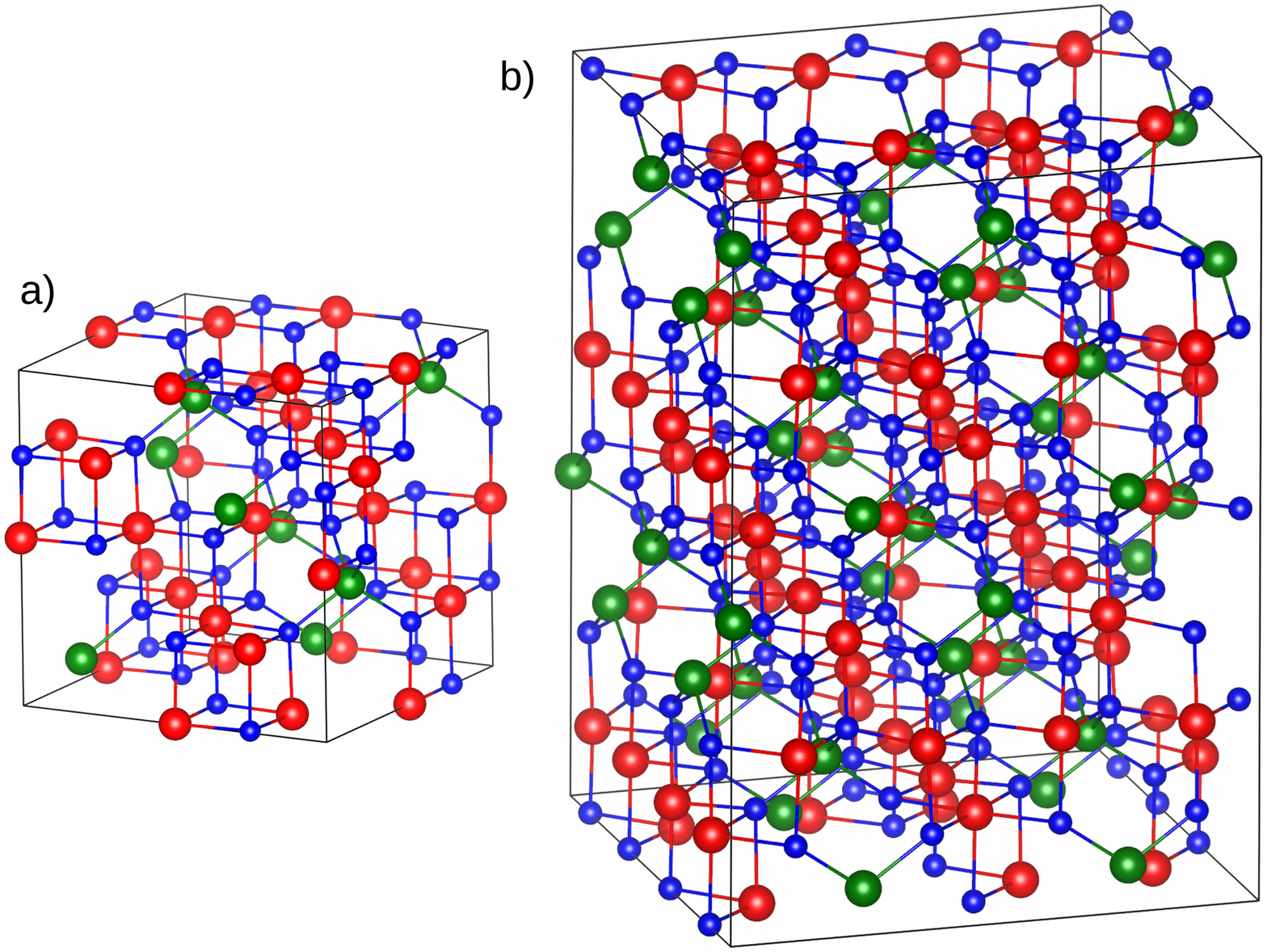}
\end{center}
\caption{The unit cells of magnetite in: 
(a) cubic $Fd\bar{3}m$ and 
(b) monoclinic $Cc$ symmetry. Green, red, and blue spheres 
represent Fe$_A$, Fe$_B$, and O atoms, respectively.}
\label{fig1}
\end{figure}

The Verwey transition in magnetite---in which the electrical 
conductivity changes discontinuously by two orders of 
magnitude~\cite{verwey1939,verwey1941}---is one of the most intriguing 
and extensively studied phenomena in condensed matter physics. The 
exceptional character of the phase transition stems from a cooperative 
mechanism involving charge, orbital, spin, and lattice degrees of 
freedom. Above the transition at $T_{\mathrm V}=124$ K, magnetite 
crystallizes in the cubic $Fd\bar{3}m$ structure \cite{bragg1915} with 
the tetrahedrally coordinated $A$-sites occupied by Fe$^{3+}$ ions and 
octahedrally coordinated $B$-sites with the mixed valency Fe$^{2.5+}$. 
At $T_{\mathrm V}$, magnetite exhibits a structural phase transition \cite{samuelsen1968,yamada1968,yoshida1977,iizumi1982,wright2001,wright2002} 
from the cubic $Fd\bar{3}m$ to the monoclinic $Cc$ symmetry 
(see Fig.~\ref{fig1}) \cite{senn2012a}.

To explain this transition, Verwey postulated a purely electronic 
mechanism of charge ordering, where the (001) planes are alternatively 
occupied by Fe$_B^{2+}$ and Fe$_B^{3+}$ ions below $T_\mathrm{V}$. 
Subsequently, this simple model was questioned on the basis of 
diffraction studies \cite{samuelsen1968,yamada1968,yoshida1977,iizumi1982} 
and a more complex charge-orbital (CO) order with 16 nonequivalent $B$ 
sites was identified below $T_\mathrm{V}$ \cite{wright2001,wright2002,senn2012a,huang2006,nazarenko2006,
schlappa2008,joly2008,lorenzo2008,blasco2011,tanaka2012,reznicek2015,reznicek2017}. 
As demonstrated by density functional theory (DFT) studies, the CO 
order is a consequence of strong local electron interactions within the 
$3d$ states and the coupling of electrons to lattice distortions
~\cite{leonov2004,jeng2004,jeng2006,piekarz2006,piekarz2007,pinto2006,yamauchi2009}.
This electron-lattice cooperative behavior results in a complicated 
pattern composed of three-site polarons, called trimerons 
\cite{senn2012a,senn2012b}. In each trimeron two end Fe$^{3+}$ cations 
are shifted towards a central Fe$^{2+}$ cation \footnote{The Fe$^{2+}$ 
denotes Fe ions with slightly smaller valency and larger occupancy of 
the minority-spin $t_{2g}$ orbitals as compared with a Fe$^{3+}$ ion}.
Recently, the soft electronic fluctuations of the trimeron order showing 
critical behavior close to the Verwey transition have been discovered
~\cite{baldini2020}. These fluctuations were described through a polaron 
tunneling model \cite{Wal02}, which involves the coupling between the 
trimeron structure and phonons.

The interaction between electronic and lattice degrees of freedom is 
reflected in the precursor effects observed above $T_{\mathrm V}$.
The critical behavior of the elastic constant $c_{44}$, which softens 
from room temperature to $T_{\mathrm V}$ \cite{moran1969,schwenk2000}, 
was explained in terms of bilinear coupling of the elastic strain to a 
fluctuation mode of the charge ordering field of $T_{2g}$ symmetry 
\cite{schwenk2000}. Critical fluctuations above $T_{\mathrm V}$ were 
investigated by diffuse scattering of electrons \cite{chiba1975}, 
neutrons \cite{fujii1975,shapiro1976,yamada1980,siratori1998}, and 
x-rays \cite{bosak2014}. Neutron measurements revealed the diffuse 
scattering at commensurate points in reciprocal space, i.e., at the 
same wave vectors as the superlattice reflections of the monoclinic 
phase, only a few kelvins above $T_{\mathrm V}$ \cite{fujii1975}.

However, even more intense maxima and in a much broader range of 
temperatures (even at room temperature) were found at incommensurate 
$\bm{q}$ vectors close to the $\Gamma$ and $X$ points of the Brillouin 
zone \cite{shapiro1976,yamada1980,siratori1998}. Shapiro \textit{et al.} 
\cite{shapiro1976} showed that neutron diffuse scattering centered at 
zero energy is coupled with transverse acoustic (TA) phonons. This 
electron-phonon coupling induces anomalous anharmonic behavior of the 
TA modes, with the largest phonon broadening found at incommensurate 
$\bm{q}$ points \cite{hoesch2013}. 

It is remarkable that an extended x-ray absorption fine structure 
(EXAFS) study \cite{subias2005} provided direct evidence that local 
distortions related to atomic displacements of the monoclinic phase are 
robust and remain unaltered during the transition. Recently, an x-ray 
pair distribution function analysis has revealed that local structural 
fluctuations persist up to the Curie transition at $T_C=850$~K 
\cite{perversi2019}. This short-range order (SRO), first postulated by 
Anderson~\cite{anderson1956}, was also identified via several probes: 
electric conductivity~\cite{ihle1986}, muon-spin-relaxation 
\cite{boekema1986}, optical conductivity~\cite{park1998}, photoemission 
\cite{schrupp2005}, magneto-optical~\cite{caicedo2014}, and resonant 
inelastic x-ray scattering (RIXS) studies~\cite{huang2017}.

All the results mentioned above indicate the important role of
the electron-phonon coupling which modifies the transport properties 
of magnetite, and may initiate the structural transition from the cubic 
to the monoclinic phase. In the theory this coupling was introduced by 
Yamada who proposed that the mechanism behind the Verwey transition 
involves the condensation of charge order fluctuations coupled to a TA 
phonon mode with $\Delta_5$ symmetry \cite{yamada1975}. Indeed, studies 
based on DFT calculations demonstrated that phonons with $\Delta_5$, 
$X_3$, and $X_4$ symmetries strongly couple to the electronic states and 
may induce the monoclinic deformation 
\cite{piekarz2006,piekarz2007,hoesch2013}. According to a detailed 
refinement, the low-temperature structure of magnetite can be described 
by a subset of phonon modes at the $\Gamma$, $\Delta$, $X$, and $W$ 
points of the Brillouin zone \cite{blasco2011,senn2013,senn2015}.

Experimentally, the phonon dispersion curves in magnetite were studied 
by inelastic neutron scattering (INS) 
\cite{watanabe1962,samuelsen1974,borroni2017b} and by inelastic x-ray 
scattering (IXS)~\cite{hoesch2013}. The Fe-projected phonon density of 
states (DOS) was measured using nuclear inelastic scattering (NIS) 
\cite{seto2003,handke2005,kolodziej2012} and detailed studies of the 
optical modes at the Brillouin zone center were performed by spontaneous 
Raman scattering and infrared reflectivity measurements~\cite{grimes1972,verble1974,degiorgi1987,graves1988,gasparov2000,gupta2002,yazdi2013,kumar2014,borroni2018,borroni2017a}.
The presence of phonon anharmonicity and strong electron-phonon coupling 
was also demonstrated by means of Raman measurements~\cite{kumar2014}.
Anomalies of the lattice vibrations that originate from strong coupling 
to electronic excitations and the spectroscopic signatures of diffusive 
modes in the electronic contribution to the Raman response function 
were observed~\cite{borroni2018}. Ultrafast broadband optical 
spectroscopy revealed that impulsive photoexcitation of particle-hole 
pairs couples to the fluctuations of the ordering field and coherently 
generates phonon modes of the ordered phase above $T_{\mathrm V}$ 
\cite{borroni2017a}. The existence of long-lived precursor fluctuations 
was further evidenced by recent INS studies~\cite{borroni2020}.

Since the lattice dynamics is strongly coupled to the electronic system 
and the SRO in the high-temperature phase of magnetite, it is important 
to explore how the phonon spectrum changes at the stuctural phase 
transition and how well these changes can be described within the DFT 
framework. To this end, in this paper, we analyze the phonon spectra 
calculated for the $Fd\bar{3}m$ cubic and the $Cc$ monoclinic structures. 
The results obtained for the monoclinic phase are in excellent agreement 
with the experimental data. In the cubic phase, the theoretical spectrum 
deviates substantially from the experimental one, which remains quite 
similar to the one in the broken symmetry phase, indicating significant 
influence of the SRO on the lattice dynamics.

The paper is organized as follows. In Sec.~\ref{cm}, the details of the 
calculation method are presented. The computed phonon dispersions and 
phonon density of states are analyzed and compared with the experimental 
data in Secs.~\ref{ps} and \ref{exp}, respectively. 
In Sec.~\ref{trimerons}, the influence of the CO order on the phonon 
modes and the mean square displacements in the monoclinic phase is studied.
The discussion and summary of the results are included in Sec.~\ref{dis}.

\section{Calculation method}
\label{cm}

The electronic and crystal structures of magnetite were optimized within 
the DFT framework implemented in the {\sc vasp} program \cite{VASP1,VASP2}.
The calculations were performed using the projector augmented-wave 
method \cite{blochl1994} and the generalized gradient approximation 
(GGA) \cite{perdew2008}. We ensured an experimentally observed 
ferrimagnetic order with the opposite orientation of magnetic moments within 
the Fe$_A$ and Fe$_B$ sublattices. Strong electronic interactions in the 
Fe($3d$) states were taken into account using the GGA+$U$ method 
\cite{lichtenstein1995} with the local Coulomb interaction parameter 
$U=4.0$~eV and Hund's exchange $J=0.8$~eV.

The cubic structure was studied in the unit cell presented in 
Fig.~\ref{fig1}(a) containing 56 atoms (4 primitive cells): 8 Fe$_A$, 
16 Fe$_B$, and 32 O atoms. The optimized lattice parameter $a=8.45$~\AA\ 
is slightly overestimated compared to the experimental value (8.39~\AA).
The crystal structure of the monoclinic $Cc$ phase was optimized within 
the unit cell containing 224 atoms [see Fig.~\ref{fig1}(b)].
The calculated lattice parameters $a=11.85$~\AA, $b=11.84$~\AA, 
$c=16.69$~\AA, and the monoclinic angle $\beta=90.20$\degree\ are in 
good agreement with the experimental data, and the obtained CO 
(trimeron) order agrees with the results presented in Refs. 
\cite{senn2012a,senn2012b}. Recently, we have used the same relaxed 
structure in the analysis of the soft charge (polaron) fluctuations 
of the trimeron order~\cite{baldini2020}.

\begin{figure}[t!]
\includegraphics[scale=0.93]{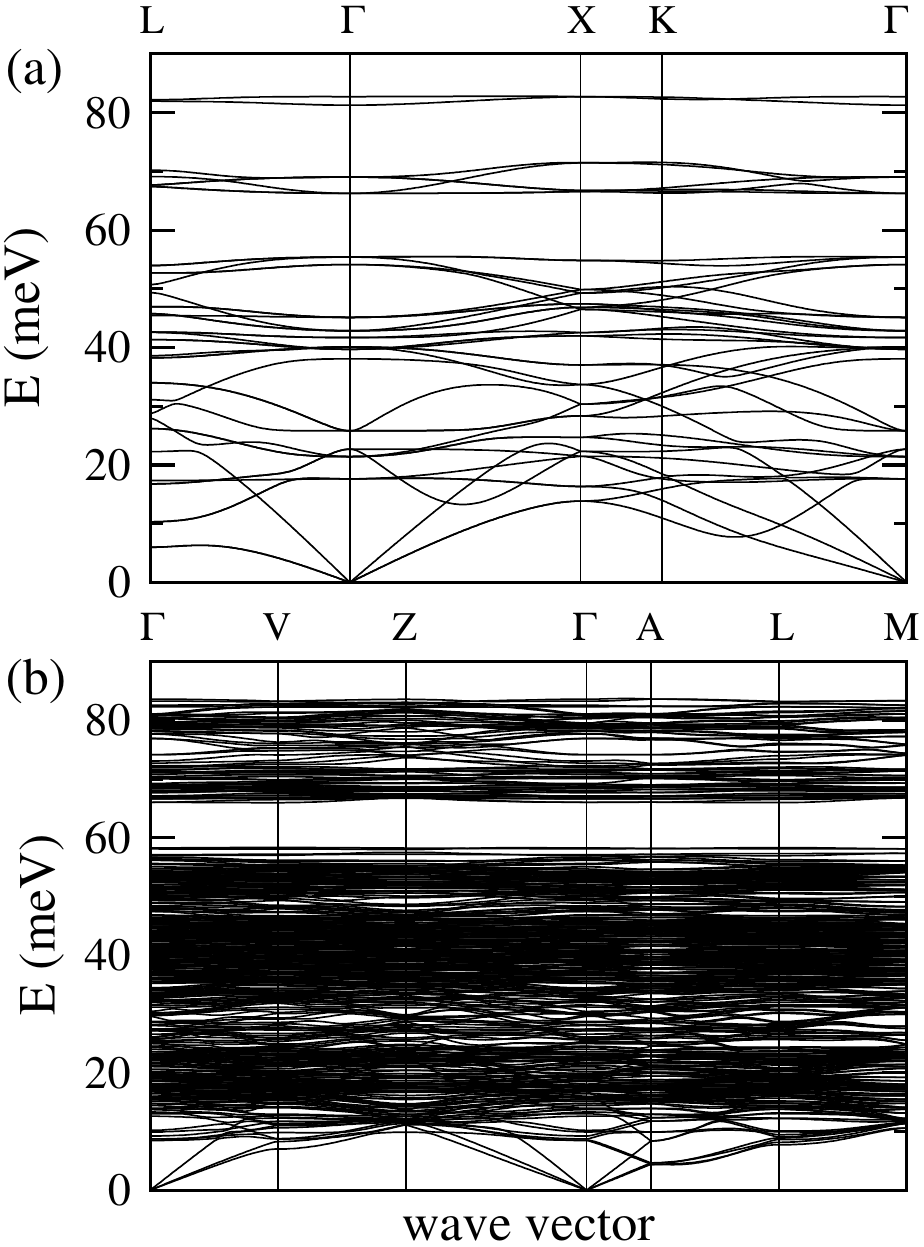}
\caption{The phonon dispersion curves of magnetite calculated for: 
(a) the cubic $Fd\bar{3}m$ and 
(b) monoclinic $Cc$ structures.}
\label{fig2}
\end{figure}

The phonon dispersion curves and phonon DOS were calculated by the 
direct method \cite{PHONON1} using the {\sc Phonon} software 
\cite{PHONON2}. The Hellmann-Feynman forces were computed through the 
displacement of all nonequivalent atoms from their equilibrium positions 
and the force-constant matrix elements were determined. By diagonalizing 
the dynamical matrix, the phonon dispersions along the high-symmetry 
directions in the first Brillouin zone were obtained. In order to 
evaluate properly the longitudinal optical-transverse optical (LO-TO) 
splitting in the monoclinic (insulating) phase, the static dielectric 
tensor and Born effective charges were determined using density 
functional perturbation theory \cite{gajdos2006}.

\section{Phonon spectra}
\label{ps}

\begin{figure}[t!]
\includegraphics[scale=0.81]{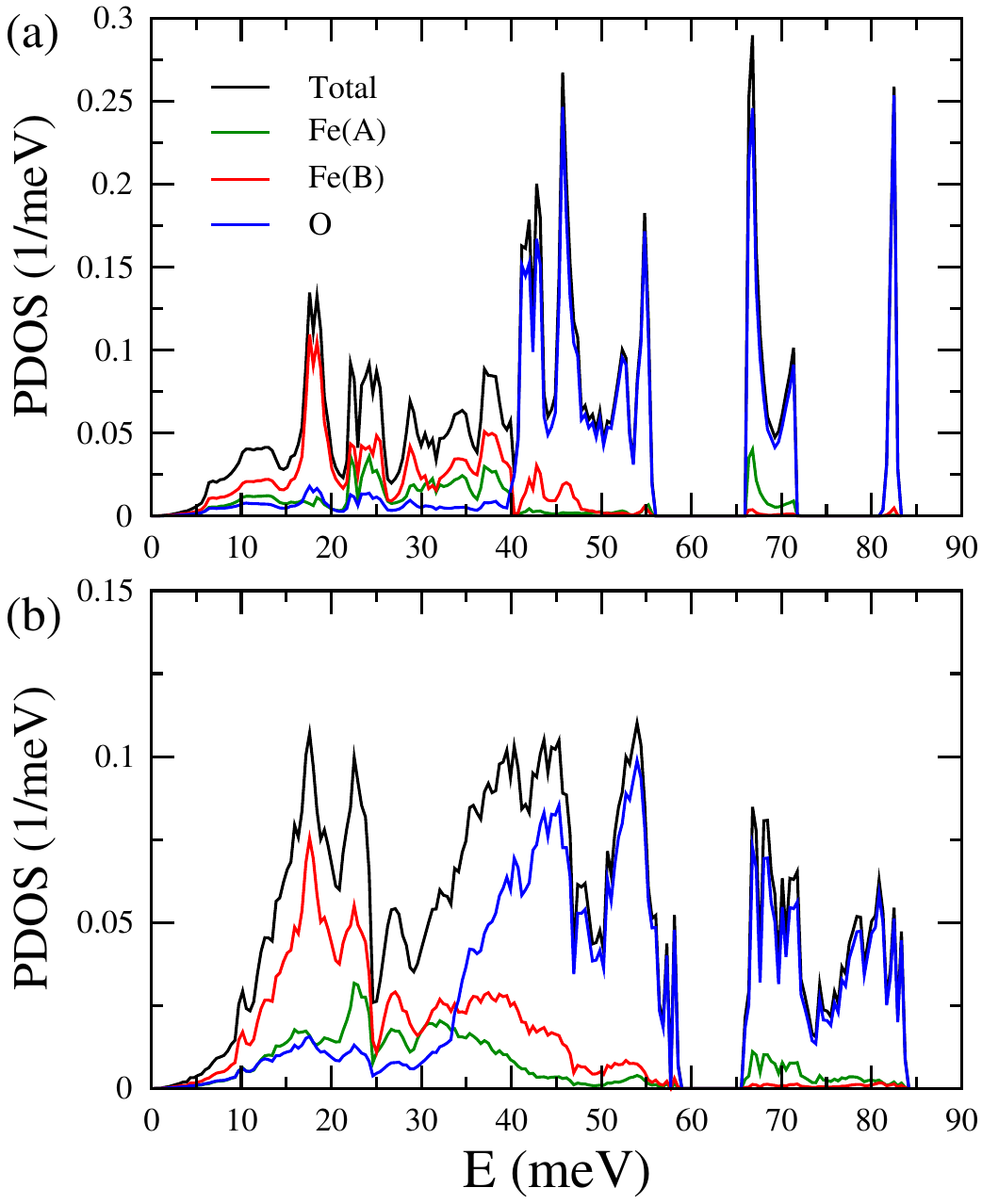}
\caption{The partial and total phonon DOS of magnetite calculated for: 
(a) the cubic $Fd\bar{3}m$ and  
(b) monoclinic $Cc$ structure.}
\label{fig3}
\end{figure}

The phonon dispersion curves and phonon DOS obtained for the cubic and 
monoclinic phases are presented in Figs.~\ref{fig2} and \ref{fig3}. In 
the cubic structure, there are 42 phonon modes. All phonon energies are 
real (i.e., both phases are stable), although some modes show anomalous 
softening close to the $L$ point, along the $\Gamma$-$X$, and the 
$K$-$\Gamma$ directions. These anomalous dispersions induce enhanced DOS 
and non-Debye behavior at the lowest energies (see Fig.~\ref{fig3}(a)). 
As we discuss later, this effect is related to the instability of the 
cubic symmetry connected with short-range distortions observed by 
diffuse scattering at temperatures above $T_{\mathrm V}$.

The low-energy range of the spectrum is dominated by vibrations of Fe 
atoms, with larger contribution of Fe atoms at the octahedral $B$ 
positions. The maximum of Fe DOS is found around 18 meV. The vibrations 
of Fe$_A$ atoms dominate at higher energies for $E\in(20,40)$ meV. 
Above 40~meV, mainly O atoms contribute to the phonon DOS, showing an 
energy cut-off at 84 meV. Two energy gaps exist between 55-66 meV and 
72-81 meV, in the cubic phase. In the band located above 65 meV, there 
is a visible contribution from Fe$_A$ atoms and a negligible 
contribution from Fe$_B$ ones.

The primitive cell of the $Cc$ structure contains 112 atoms. Therefore, the 
number of phonon modes increases to 336. The partial DOS of Fe atoms is 
increased above 50 meV compared to the cubic phase and they contribute 
to the highest modes up to 80 meV. The lower gap is reduced to about 7 
meV and the higher one is completely closed.

\begin{figure}[b!]
\includegraphics[scale=0.31]{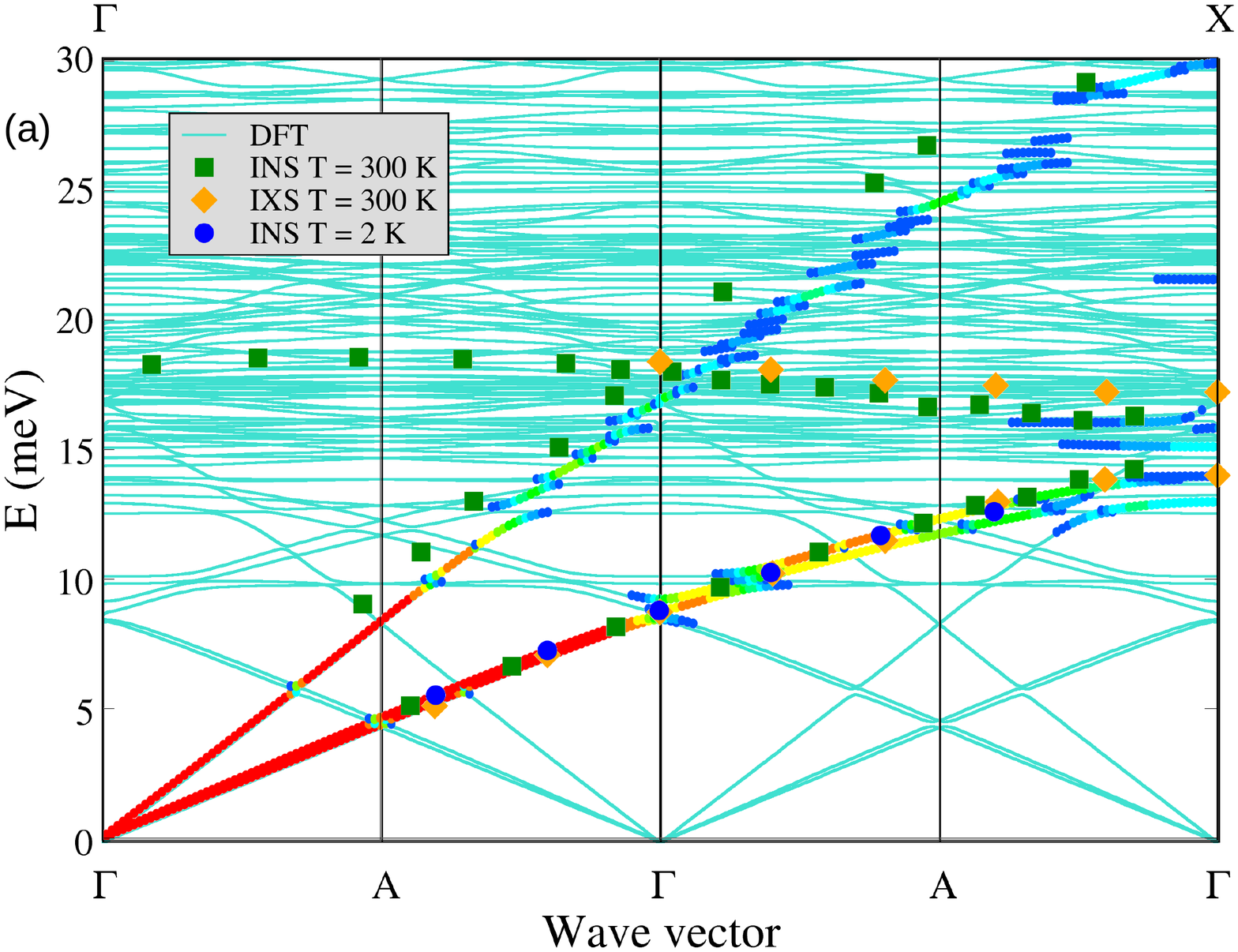}
\vskip -.3cm
\includegraphics[scale=0.31]{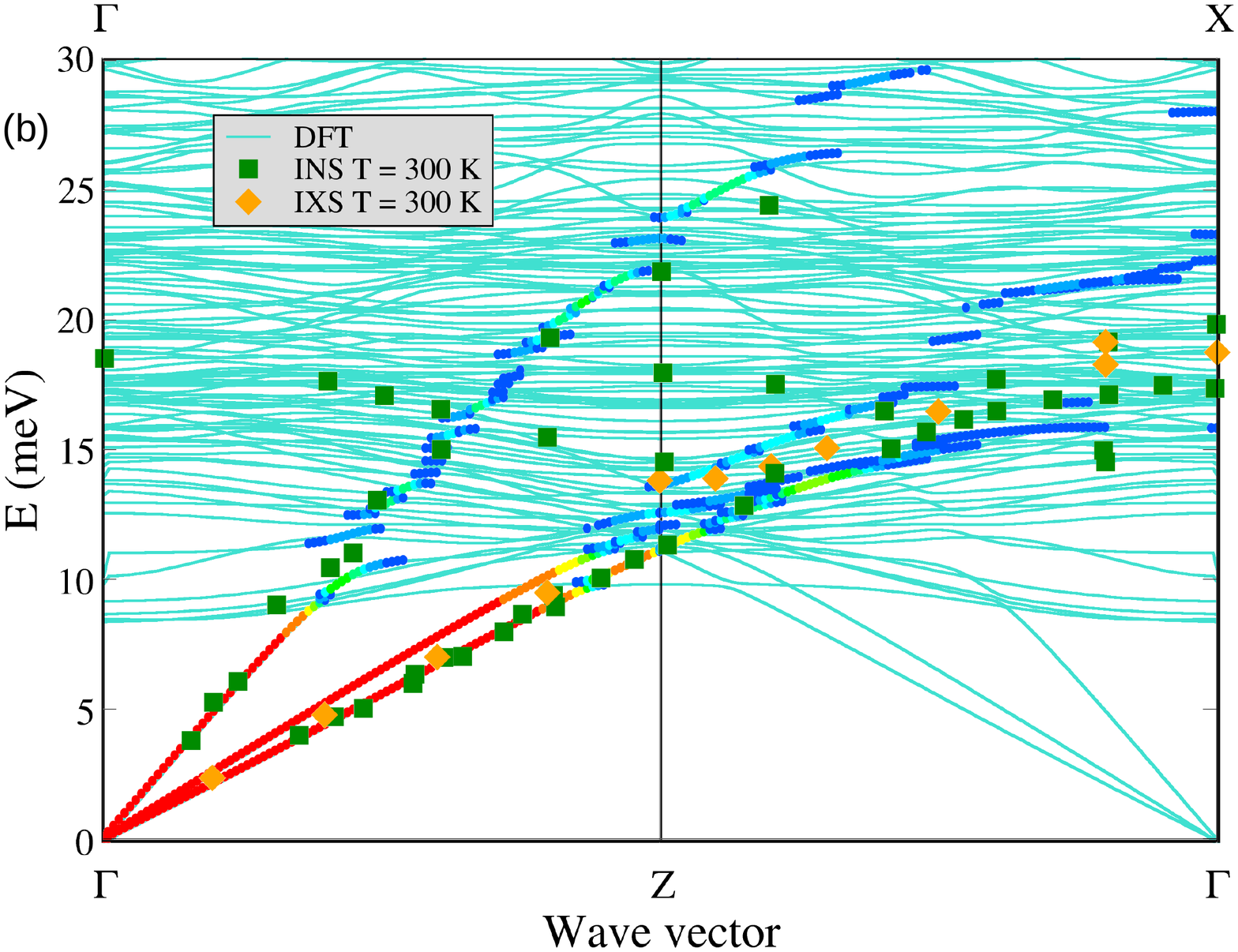}
\caption{The phonon dispersion curves calculated for the $Cc$ structure compared 
with INS experimental data at $T=2$ K taken from Refs.
\cite{samuelsen1974} (green squares) and \cite{borroni2017b} (blue 
circles) as well as IXS data \cite{hoesch2013} (orange diamonds) along: 
(a) the $\Delta$[001] and 
(b) $\Sigma$[110] directions of the cubic Brillouin zone. 
The phonon energies with the dynamical form factor above a certain
threshold $t$ are plotted in blue ($t=10\%$), green ($t=50\%$), yellow ($t=70\%$), and red ($t=90\%$).
The high symmetry points of the cubic (monoclinic) Brillouin zone are 
presented in the top (bottom) of the figures.}
\label{fig4}
\end{figure}

\section{Comparison with experimental data}
\label{exp}

In Fig.~\ref{fig4}, the phonon dispersions obtained for the $Cc$ 
structure are compared with the available experimental data. The 
dispersions measured above $T_{\mathrm V}$ by INS~\cite{samuelsen1974} 
and IXS~\cite{hoesch2013} are presented in the extended Brillouin zone 
of the monoclinic structure. The INS measurements at $T=2$~K, well below 
the Verwey transition, were performed along the $\Gamma$-$A$ direction 
of the monoclinic Brillouin zone~\cite{borroni2017a}. The dispersions 
are presented for the \mbox{$\Gamma$-$A$} [Fig.~\ref{fig4}(a)] and 
$\Gamma$-$Z$ [Fig.~\ref{fig4}(b)] directions, which correspond to the 
$\Gamma$-$X$ $\Delta$[0,0,1] and $\Gamma$-$K$-$X$ $\Sigma$[1,1,0] 
directions in the cubic symmetry, respectively. The lowest optical modes 
at the $\Gamma$ point around 8 meV result from the splitting and folding 
of the TA branch with $\Delta_5$ symmetry in the cubic structure. 

Additionally, in Fig.~\ref{fig4}, we plot the formfactor, which 
describes the intensity (strength) of phonon modes
\begin{equation}
F(\bm{k},j)=\left|\,\sum_{\alpha}
\frac{\bm{e}(\bm{k},j;\alpha)}{\sqrt{M_{\alpha}}}\,\right|^2, 
\end{equation}
where $\bm{e}(\bm{k},j;\alpha)$ is the polarization vector, $\bm{k}$ is the phonon wave vector, $j$ denotes the phonon branch, 
and $M_{\alpha}$ is the mass of atom $\alpha$. It can be observed that 
the phonon modes with the strongest intensity in the monoclinic 
structure correspond to the acoustic branches of the cubic structure.

\begin{figure}[b!]
\includegraphics[scale=0.31]{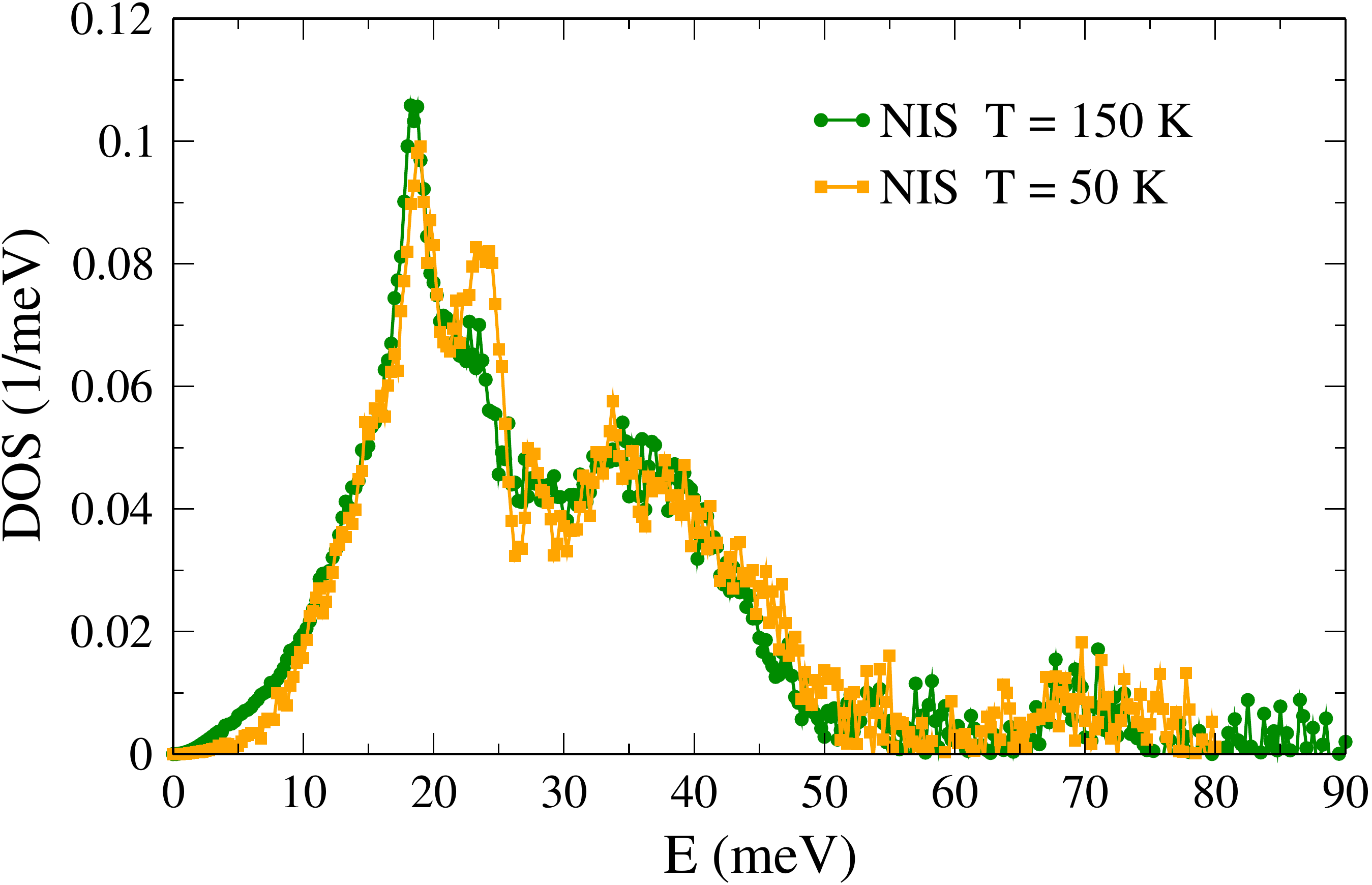}
\caption{The Fe-partial phonon DOS in magnetite measured by NIS 
at 50 K and 150 K taken from Ref.~\cite{kolodziej2012}.}
\label{fig5}
\end{figure}

A good agreement between the phonon energies measured below and above 
$T_{\mathrm V}$ with the theoretical dispersions calculated for the 
low-temperature $Cc$ structure demonstrates that phonon energies in 
magnetite depend weakly on temperature. It also demonstrates that the 
SRO existing above $T_{\mathrm V}$ preserves the local geometry of the 
$Cc$ structure. The lowest TA dispersion measured below $T_{\mathrm V}$ 
shows slightly higher values than the corresponding branch in the cubic 
phase \cite{borroni2017a}. We remark that the phonon dispersions 
calculated for the cubic phase were compared with the experimental 
points measured above $T_{\mathrm V}$ in the previous studies 
\cite{piekarz2006,piekarz2007,hoesch2013}.

The influence of the phase transition on the partial Fe phonon DOS was 
studied by NIS measurements~\cite{handke2005,kolodziej2012}. The NIS 
spectra obtained at temperatures $T=50$ K and 150 K~\cite{kolodziej2012} 
are compared in Fig.~\ref{fig5}. The main differences are observed at 
the lowest energies below 10 meV and around two peaks at 24 meV and 27 
meV. The main peak at 18 meV is weakly affected by the transition and 
is only slightly shifted to higher energies below the Verwey transition.
The increase of phonon DOS in the cubic phase below 10 meV is related 
to the quasielastic (diffuse) scattering observed at low energies above 
the Verwey transition. The stiffening of the monoclinic lattice is 
evidenced by the increase of the Lamb-M\"{o}ssbauer factor below 
$T_{\mathrm V}$~\cite{handke2005} and is consistent with the INS 
measurements~\cite{borroni2017a}. 
We calculated the first moment of the phonon DOS, which gives the average value of the
phonon energy. We found that its value increases, when going from the cubic (42.09~meV)
to the monoclinic (42.22~meV) phase. If we take only the contribution from the Fe$_B$ atoms,
this change is more pronounced (from 26.74~meV to 27.32~meV).  
As we discuss later, the peak at 24 meV is strongly influenced by the CO order in the $Cc$ structure and 
this effect may explain its modification by the Verwey transition.

\begin{figure}[t!]
\includegraphics[scale=0.91]{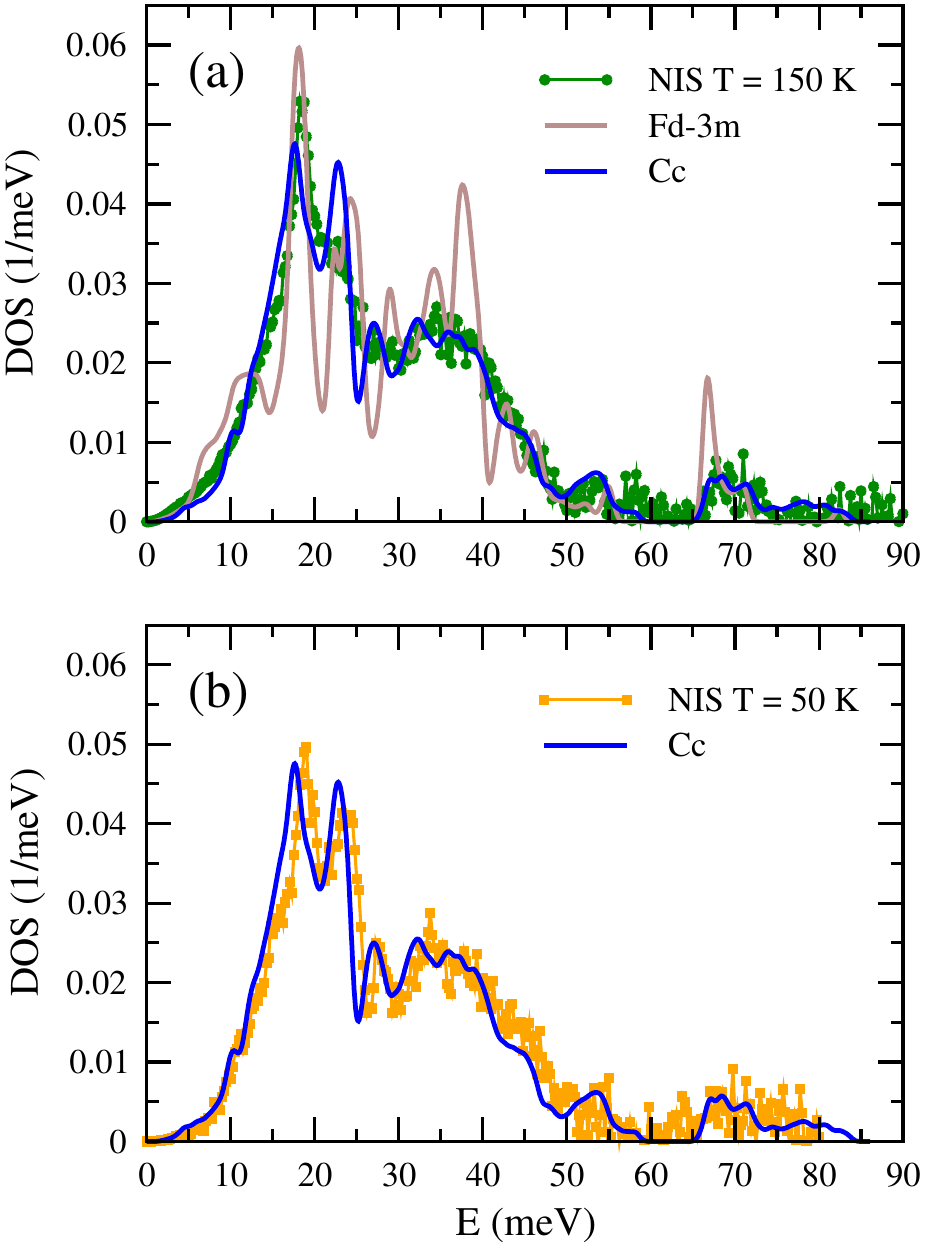}
\caption{The Fe-partial phonon DOS in magnetite measured at: 
(a) $T=150$~K and calculated for the cubic 
$Fd\bar{3}m$ and monoclinic $Cc$ structure and 
(b) $T=50$~K and calculated for the monoclinic $Cc$ 
symmetry. The experimental data are taken from 
Ref.~\cite{kolodziej2012}.}
\label{fig6}
\end{figure}

The Fe-projected phonon DOS calculated for the cubic and monoclinic 
symmetries is compared with the experimental data obtained above and 
below $T_{\mathrm V}$ in Fig.~\ref{fig6}. All theoretical spectra were 
broadened by a Gaussian with a width of 1 meV, which is a typical 
experimental resolution. In the phonon DOS calculated for the cubic 
structure, the main peaks coincide with the high intensity DOS measured 
by NIS. However, apart from the highest peak at 18 meV, all other 
spectral features are much broader than the theoretical peaks.
In addition, one finds a broad peak centered at 12 meV not observed in 
the experimental spectrum. This additional enhancement of phonon DOS 
results from the anomalous dispersions found in the cubic phase 
(see Fig.~\ref{fig2}).

The phonon DOS calculated for the monoclinic structure shows much better 
agreement with the experimental data, even for the spectrum obtained above 
$T_{\mathrm V}$. Apart from a small shift of the theoretical energies to 
lower energies, the positions and widths of all peaks are very well 
reproduced. The intensity of the lowest peak around 18 meV is slightly 
underestimated, while the higher peak around 24 meV is overestimated 
when compared to the experimental DOS. The position and intensity of the 
peak around 27 meV agree very well with the experiment. 

We emphasize that the theory captures the essential features of the 
phonon DOS at higher energies which exhibits most of the features 
obtained by the NIS measurements. Such good agreement demonstrates that 
the electronic ground state and interatomic forces are very well 
described by the DFT calculations. It also shows that all atomic 
positions and their valence states are well defined in the 
low-temperature monoclinic phase unlike in the high-temperature phase, 
where the SRO, present in reality, is not modeled by theoretical 
calculations. We note that the phonon DOS calculated in the $Cc$ 
structure agrees much better with the high-temperature data than the 
computation in the cubic structure [see Fig.~\ref{fig6}(a)]. This difference between the 
low-temperature and high-temperature structures highlights again the 
importance of short range correlations above $T_{\mathrm V}$.

\section{The impact of trimeron order on phonons}
\label{trimerons}

\begin{center}
\begin{figure}[t!]
\includegraphics[scale=0.62]{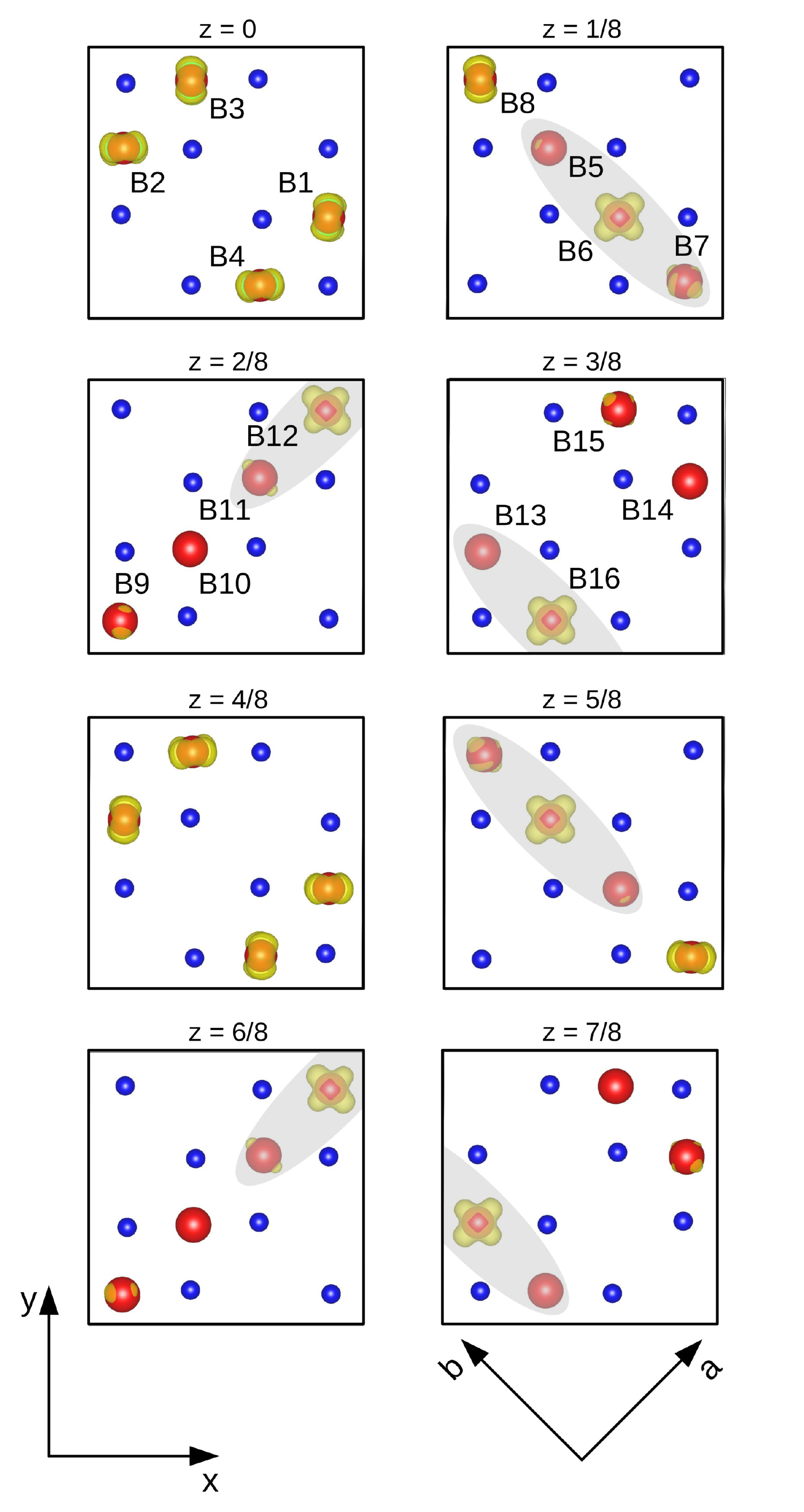}
\caption{The charge-orbital order in the $Cc$ structure presented 
for each $(0, 0, z)$ plane separately. The trimerons related to the 
occupied $d_{xy}$ and lying in the $(a,b)$ plane are shown as gray 
ovals. 
The $B$ sites are denoted as in Refs.~\cite{jeng2006,yamauchi2009}.}
\label{fig7}
\end{figure}
\end{center}

In the $Cc$ structure, there are 16 nonequivalent $B$ positions. Half of 
the Fe$_B$ ions have a nominal $2+$ valence state and half of them are 
$3+$. The CO ordering for each Fe$_B$-O plane (defined here by the 
fractional coordinate $z$) of the primitive $Cc$ cell is presented in 
Fig. \ref{fig7}. The main crystallographic directions $a$ and $b$ of the 
$Cc$ unit cell (see Fig.~\ref{fig1}) are rotated by 45\degree\ with 
respect to the $x$ and $y$ axes defined by the main directions of the 
cubic phase (both presented at the bottom of the figure). In each plane, 
the Fe$_B$ atoms form one-dimensional chains with the smallest nearest 
neighbor distances along the $a$ or $b$ direction. In the $z=0$ and 
$z=4/8$ planes, there are only Fe$^{2+}$ ions with occupied $d_{xz}$ 
and $d_{yz}$ orbitals, alternating along the $a$ direction ($B1$, $B2$, 
$B3$, and $B4$ sites). The $z=1/8$ or $z=5/8$ planes contain each two 
Fe$^{2+}$ ions with $d_{xy}$ ($B6$) and $d_{yz}$ or $d_{xz}$ orbitals 
($B8$), while only one Fe$^{2+}$ ion with the $d_{xy}$ orbital is 
located in the $z=2/8$ ($B12$) and $z=3/8$ ($B16$) planes.

The Fe$_B^{2+}$ ions are the centers of trimerons, which are oriented 
according to the symmetries of the occupied orbitals. In Fig.~\ref{fig7}, 
the trimerons with occupied $d_{xy}$ orbitals and oriented along the $a$ 
or $b$ directions are depicted by the gray ovals. The other trimerons 
centered at the $B$ sites located in the planes with $z=0$, $1/8$, 
$4/8$, and $5/8$ are aligned along the diagonal $xz$ or $yz$ direction 
(not shown in Fig.~\ref{fig7}). The trimerons are not isolated objects; 
most of them are connected and have common Fe$_B^{3+}$ ions at their 
ends. In two cases, the end of one trimeron coincides with the center of 
the other one located in the $z=0$ or $z=4/8$ plane. A detailed analysis 
of the trimeron order is presented in~Refs.~\cite{senn2012a,senn2012b}.

\begin{figure}[t!]
\includegraphics[scale=0.81]{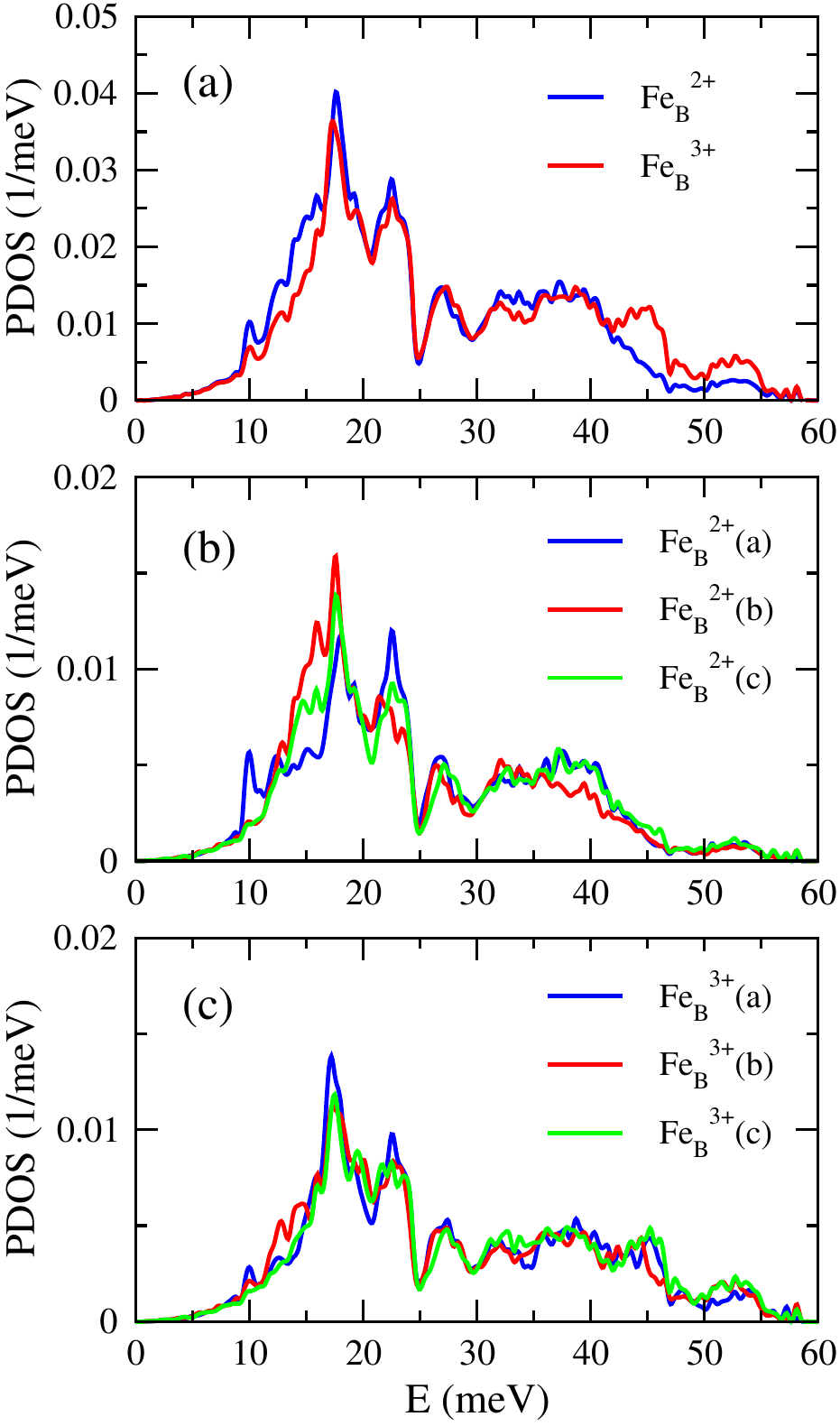}
\caption{(a) The partial phonon DOS of Fe$_B^{2+}$ and Fe$_B^{3+}$ ions and
the direction-projected phonon DOS of (b) Fe$^{2+}_B$ and (c) Fe$^{3+}_B$ ions in the mono\-clinic 
$Cc$ structure.}
\label{fig8}
\end{figure}

The CO distribution in the $Cc$ structure affects the interatomic forces 
and therefore also influences the phonon energies. This effect can be 
clearly observed in Fig.~\ref{fig8}(a), where the phonon DOS is plotted 
for the Fe$_B^{2+}$ and Fe$_B^{3+}$ cations separately. The partial DOS 
of Fe$_B^{2+}$ ions show higher values at lower energies and the main 
peaks below 24 meV are larger than for the Fe$_B^{3+}$ ions. 
In contrast, at higher energies (above 40 meV), the phonon DOS is 
dominated by the Fe$_B^{3+}$ vibrations. It can be related to an enhanced  
Coulomb interaction between the Fe cations with the larger valence state 
and the oxygen anions.

The charge distribution in the orbitals of Fe ions at the $B$ sites is 
also highly anisotropic. If we consider two crystallographic directions 
$a$ and $b$ in the monoclinic $Cc$ structure, the differences result 
from the distribution of charges in the $z=0$ and $z=4/8$ planes, in 
which the chains of Fe$_B^{2+}$ ions are oriented along the $a$ 
direction, and the $z=1/8$ and $z=5/8$ planes containing two Fe$^{2+}$ 
ions in the chains oriented along the $b$ direction. This anisotropy in 
charge ordering is related to slightly different values of the lattice 
parameters along the $a$ and $b$ directions. Therefore, we should also 
see a variation in the phonon spectra corresponding to these two 
perpendicular directions. 

This expectation is confirmed by the data displayed in 
Fig.~\ref{fig8}(b) and Fig.~\ref{fig8}(c), where we present the phonon DOS 
of Fe$^{2+}_B$ and Fe$^{3+}_B$ vibrations, respectively, along 
three nonequivalent crystallographic directions of the $Cc$ unit cell. 
Indeed, for these three directions, the phonon DOS exhibits 
differences which are more significant for the Fe$^{2+}_B$ ions.
It shows that the phonon anisotropy results mainly from the orbital ordering
of the spin-down states at the Fe$^{2+}_B$ sites.
The peak observed at 10 meV is mainly related to the vibrations along the $a$ axis. 
The phonon DOS above 10 meV and the highest peak around 18 meV show a 
the anisotropy with the largest contribution from the $b$ component. 
The peak around 24 meV exhibits even larger anisotropy and is dominated by 
the vibrations along the $a$ axis. 

\begin{figure}[t!]
\includegraphics[scale=0.8]{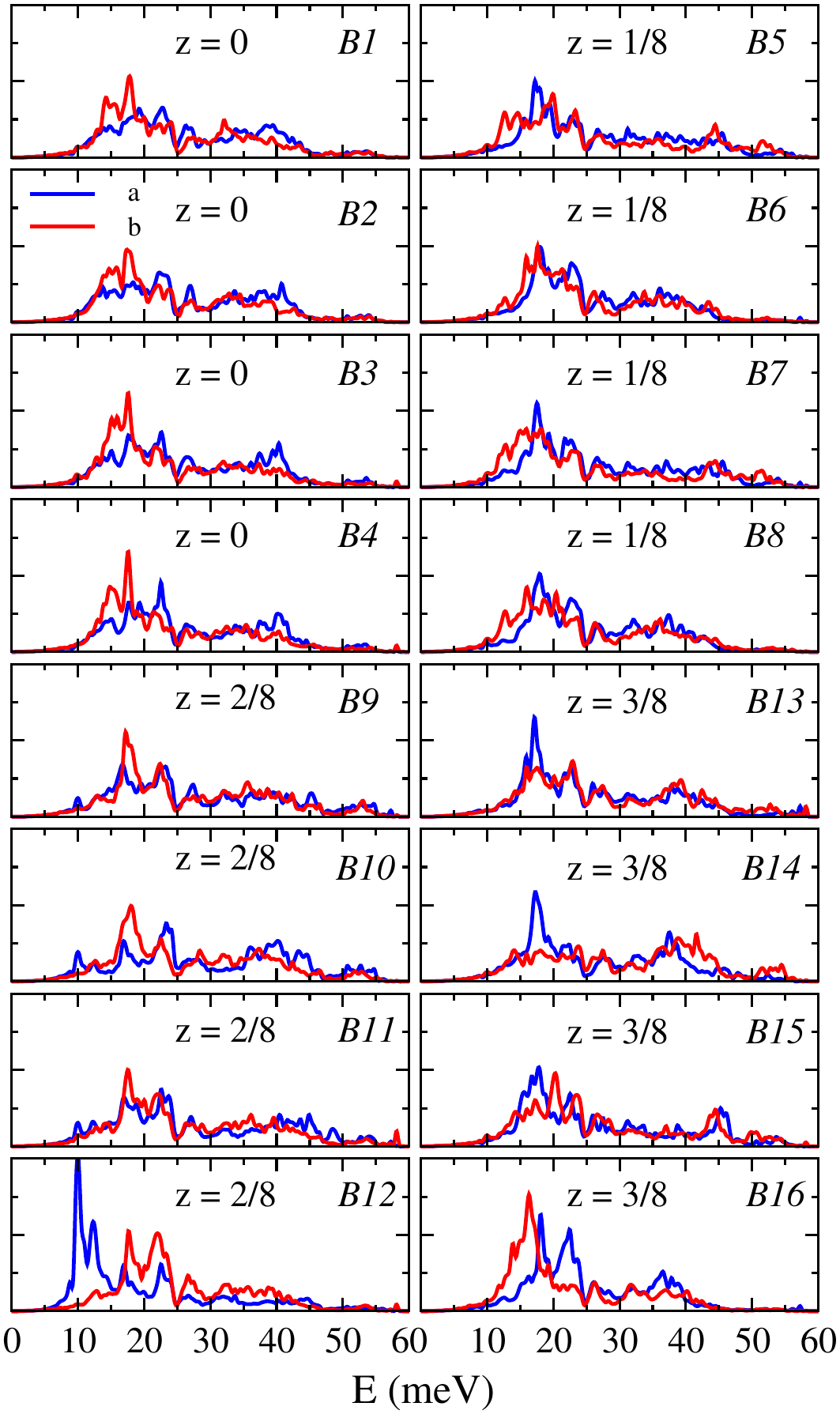}
\caption{The phonon DOS projected into all nonequivalent Fe$_B$ ions and 
two crystallographic directions $a$ and $b$ in the monoclinic structure.}
\label{fig9}
\end{figure}

The influence of the CO anisotropy observed here may explain the 
significant differences between the NIS spectra measured in the cubic 
and monoclinic phases around 24 meV. One finds that this peak is 
strongly suppressed in the cubic structure, which can be related to the 
``melting" of the trimeron order. We expect that the average of the two 
directions ($a$ and $b$) provides a good model of the dynamical 
structure factor in the cubic phase. However, this produces too much 
weight in the 24 meV peak. As less stable trimeron configurations are 
populated thermally, we expect some phonons to soften. This behavior may 
explain the transfer of spectral weight from the 24 meV peak to the 18 
meV peak. Irrespective of this feature, we propose that the relative 
weight of these peaks can be used to measure the degree of SRO.

\begin{table}[!t]
\caption{Mean square displacements of Fe$_B$ ions (in \AA$^2$) 
calculated at $T=100$ K and their nominal valence states.}
\begin{ruledtabular}
\begin{tabular}{ccccc}
$B$ site & Valence & $\langle U^2_{a}\rangle$ & $\langle U^2_{b}\rangle$ & $\langle U^2_{c}\rangle$ \\
\hline
$B1$    &  2+  &  0.002307  &  0.002595 & 0.002550  \\
$B2$    &  2+  &  0.002323  & 0.002589  & 0.002677  \\
$B3$    &  2+  &  0.002301  & 0.002674  & 0.002593  \\
$B4$    &  2+  &  0.002266  & 0.002638  & 0.002607  \\
$B5$    &  3+  &  0.002271  & 0.002588  & 0.002232  \\
$B6$    &  2+  &  0.002257  & 0.002485  & 0.002308  \\
$B7$    &  3+  &  0.002229  & 0.002710  & 0.002246  \\
$B8$    &  2+  &  0.002272  & 0.002633  & 0.002289  \\
$B9$    &  3+  &  0.002305  & 0.002309  & 0.002171  \\
$B10$   &  3+  &  0.002231  & 0.002335  & 0.002216  \\
$B11$   &  3+  &  0.002343  & 0.002284  & 0.002191  \\
$B12$   &  2+  &  0.004078  & 0.002326  & 0.002340  \\
$B13$   &  3+  &  0.002340  & 0.002256  & 0.002207  \\
$B14$   &  3+  &  0.002298  & 0.002098  & 0.002149  \\
$B15$   &  3+  &  0.002299  & 0.002333  & 0.002192  \\
$B16$   &  2+  &  0.002293  & 0.002945  & 0.002227
\end{tabular}
\end{ruledtabular}
\label{tab_msd}
\end{table}

To better understand the origin of phonon anisotropy between the $a$ 
and $b$ directions, we analyze the phonon DOS along these two directions 
for each nonequivalent Fe$_B$ ion. We choose 16 $B$ sites belonging to 
the four lowest planes with $z=0$, 1/8, 2/8, and 3/8 
(see Fig.~\ref{fig7}) and present the corresponding phonon DOSs in 
Fig.~\ref{fig9}. In the $z=0$ plane, there are four Fe$_B^{2+}$ ions 
showing similar DOS. There is a visible anisotropy with the larger 
contribution of vibrations along the $b$ direction at lower energies. 
This is caused by the larger Fe-Fe distances and smaller interatomic 
forces along the direction perpendicular to the $B$-chains. In the 
$z=1/8$ plane, there is some difference between the ions with 
different valence, with slightly higher anisotropy found for the 
Fe$^{3+}$ ions ($B5$ and $B7$). 

The largest differences are found in 
the two other planes, each containing only one Fe$_B^{2+}$ ion with an 
occupied $d_{xy}$ orbital ($B12$ and $B16$). The DOS at the $B12$ site 
along the $a$ direction is strongly shifted to lower energies and 
exhibits a large peak at 10 meV. It shows the greatest shift to lower 
energies compared to the other Fe$_B$ ions. Interestingly, these 
vibrations are along the $B$-chains. Similarly, the phonon DOS 
projected onto the $B16$ site shows a large anisotropy. Also in this 
case, the energies of vibrations along the $B$-chains, which in the 
$z=3/8$ plane are parallel to the $b$ direction, are shifted to lower 
values with a maximum around 15 meV. These two peaks at 10 and 15 meV 
come only from two $B$ sites, and thus their contributions to the total 
DOS are not very strong. However, they are clearly visible as shoulders 
in the experimental and theoretical DOSs presented in Fig.~\ref{fig6}.

The dynamical anisotropy was further investigated by calculating the 
mean square displacement (MSD) tensor. The diagonal elements of the MSD 
tensors of all inequivalent Fe$_B$ atoms obtained for the monoclinic 
structure at $T=100$ K are presented in Tab.~\ref{tab_msd}. The largest 
value of the MSD is found for the $B12$ Fe atom, which shows the 
strongest vibrational anisotropy in the phonon DOS. The average 
displacement of this Fe atom along the $a$ direction exceeds 70\% and is
larger than along the $b$ direction. The second largest MSD is for the 
$B16$ atom along the $b$ direction. These enhanced atomic displacements 
should couple strongly to charge fluctuations between Fe$_B$ ions and 
participate in the polaronic soft modes recently observed in optical 
experiments~\cite{baldini2020}.

\section{Final discussion and summary}
\label{dis}

The results presented in the previous section demonstrate a significant 
impact of the charge-orbital order on lattice dynamics in the monoclinic 
phase of magnetite. A very good agreement between the calculated and 
measured phonon spectra indicates that the coupling between the 
charge-orbital distribution (trimerons) and phonons is well described in 
our calculations. Through a detailed analysis of the phonon DOS 
projected into all nonequivalent $B$ sites, we revealed that some 
trimerons modify strongly the vibrations of central Fe$_B^{2+}$ ions. 
This trimeron-phonon coupling is especially strong for the trimerons 
oriented along the main directions of the monoclinic $Cc$ structure 
inducing a large shift of phonon DOS to lower energies.
Our results show for the first time that the charge-orbital ordering 
induce strong anisotropy of the lattice dynamics in the monoclinic
phase of magnetite.  
 
The obtained results can be compared with other studies which probe the 
local charge distribution on Fe$_B$ ions. The NMR and M\"{o}ssbauer 
measurements combined with DFT calculations revealed that all 
$B$-sites can be divided into three groups with the ratio 8:5:3 
\cite{reznicek2015,reznicek2017,chlan2018,kolodziej2020}. The first group consists of 8 
Fe$^{3+}$ ions, the second group includes 5 Fe$^{2+}$ ions in which 
electrons occupy the $d_{xz}$ or $d_{yz}$ orbitals, and the third group 
comprises 3 Fe$^{2+}$ ions with occupied $d_{xy}$ orbitals. The last 
group of Fe ions is characterized by lower effective magnetic fields and 
larger electric field gradients~\cite{reznicek2017}. Interestingly, two 
$B$-sites that show the strongest phonon anisotropy belong to this group 
of ions ($B12$ and $B16$ sites, which correspond to $B7$ and $B16$ sites 
in Refs.~\cite{reznicek2015,reznicek2017}). However, a phonon DOS 
projected into a single site depends not only on the local valence but 
also on the charge distribution in the neighborhood of a given atom.
This explains the different phonon DOS of the $B6$ site, which also 
belongs to the third group, when compared to the $B12$ and $B16$ sites.

The present studies demonstrate the influence of static charge-orbital 
order on phonons. We remark that recent optical conductivity and 
pump-probe experiments discovered soft electronic modes, which can be 
explained as the excitations of the trimeron order coupled to atomic 
vibrations~\cite{baldini2020}. To model these modes, we 
considered the coherent tunneling of polarons, which corresponds to 
charge fluctuations between the $B12$ and $B9$ sites in the $z=2/8$ 
plane. Such tunneling induces effectively a shift of the trimeron along 
the $a$ direction. Similar movement is possible along the $b$ direction 
in the $z=3/8$ plane, where the charge localized at the $B16$ site can 
tunnel to the $B13$ or $B15$ site. The present calculations show that 
the trimerons in these two planes couple strongly to the low-energy 
vibrations with the largest displacements, and therefore they should 
participate in the polaronic excitations discovered recently by 
terahertz spectroscopy~\cite{baldini2020}. These excitations---which 
propagate along two perpendicular directions---should have different 
energies due to the anisotropy in lattice and electronic dynamics. 
This hypothesis can be confirmed in the future by means of 
polarization-resolved terahertz experiments on untwinned crystals.

We have shown that
the existence of the short-range order above the Verwey transition, 
preserving the local geometry of the monoclinic structure, is well 
supported by various experiments~\cite{subias2005,perversi2019}. Since 
the nearest neighbor distances have a dominant impact on interatomic 
forces and phonon energies, the presence of the short-range order may 
explain small differences between the phonon spectra measured below and 
above $T_\mathbf{V}$. Moreover, the short-range order explains why the 
phonon DOS calculated for the $Cc$ structure agrees better with the 
high-temperature NIS data than the computation performed for the perfect 
cubic structure. Since the deviation from the ideal cubic geometry is 
observed up to the Curie temperature \cite{perversi2019}, the 
short-range order may determine the lattice dynamics and the 
thermodynamic properties of magnetite in a wide range of temperatures.

We suggest that the influence of the static charge-orbital order on 
phonons found in our studies could still be present above $T_V$. 
However, dynamical effects become very important in the high-temperature 
phase. The existence of charge fluctuations coupled to phonons is the 
origin of strong neutron and x-ray diffuse scattering 
\cite{fujii1975,shapiro1976,yamada1980,siratori1998,bosak2014} as well 
as the anharmonic effects revealed by the IXS studies~\cite{hoesch2013}.
In this case, both the static displacements from the high-symmetry 
cubic positions and charge fluctuations may contribute to anharmonic 
behavior above the Verwey transition.

In summary, we have investigated the lattice dynamical properties of 
magnetite by performing the DFT calculations for the cubic and 
monoclinic structures. The results obtained for the low-temperature 
monoclinic phase show a very good agreement with the experimental data, 
indicating that the electronic ground state and interatomic forces are 
well reproduced in the calculations. Most importantly, we have 
demonstrated that the trimeron order strongly influences the vibrational 
properties of Fe atoms and this we take as an evidence for strong 
trimeron-phonon coupling.

In particular, the trimerons with occupied 
$d_{xy}$ orbitals induce a strong dynamical anisotropy between the main 
crystallographic directions in the monoclinic structure. 
This anisotropy is related to the enhanced atomic displacements, which
may be closely related to the soft charge modes that were observed 
in the optical measurements~\cite{baldini2020}.
The phonon density of states calculated for the perfect 
cubic structure deviates from that of the NIS measurements, which can be 
explained by the existence of the short-range order in magnetite above 
the Verwey transition. The local lattice distortions, which modify the 
charge distribution, influence the interatomic forces and determine the 
phonon energies at higher temperatures. 

\subsection*{Acknowledgements}

This work was supported by the European Regional Development Fund in
the IT4Innovations National Supercomputing center---Path to Exascale
project, No. CZ.02.1.01/0.0/0.0/16013/0001791 within the Operational
Programme Research, Development and Education, and Donau project
No. 8X20050 by Ministry and Education of the Czech Republic.
P.P. and A.M.O. kindly acknowledge the support by Narodowe Centrum Nauki (NCN, Poland), 
Projects No. 2017/25/B/ST3/02586 and No. 2016/23/B/ST3/00839. 
A. M. Ole\'s is grateful for the Alexander von Humboldt Foundation 
Fellowship \mbox{(Humboldt-Forschungspreis)}. 
W.T. acknowledges support from the Polish National Agency for Academic Exchange under 
the “Polish Returns 2019” Program, Grant No. PPN/PPO/2019/1/00014/U/0001.
Work at MIT was supported by the US Department of Energy, BES DMSE, Award number 
DE-FG02-08ER46521 and by the Gordon and Betty Moore Foundation's EPiQS 
Initiative grant GBMF9459. E.B. and C.A.B. acknowledge additional 
support from the Swiss National Science Foundation under Fellowships 
P2ELP2-172290 and P400P2-183842 and the National Science Foundation 
Graduate Research Fellowship under Grant No. 1122374, respectively.

\bibliography{magnetite}

\end{document}